\documentstyle[preprint,aps,pre,eqsecnum,epsfig]{revtex}
\draft
\begin{document}

\title{Tricritical Points in the Sherrington-Kirkpatrick Model in the
Presence of Discrete Random Fields}

\vskip \baselineskip

\author{Jo\~{a}o M. de Ara\'{u}jo,}

\address{
Departamento de Ci\^{e}ncias Naturais \\
Universidade Estadual do Rio Grande do Norte \\
59610-210 \hspace{8mm} Mossor\'o - RN \hspace{8mm} Brazil}

\author{Fernando D. Nobre, and Francisco A. da Costa}

\address{
Departamento de F\'{\i}sica Te\'orica e Experimental \\
Universidade Federal do Rio Grande do Norte \\
Campus Universit\'{a}rio -- Caixa Postal 1641 \\
59072-970 \hspace{8mm} Natal - RN \hspace{8mm} Brazil}

\date{\today}
\maketitle

\newpage

\begin{abstract}
The infinite-range-interaction Ising spin glass is considered in the presence
of an external random magnetic field following a trimodal (three-peak)
distribution. Such a distribution corresponds to a bimodal added to a
probability $p_{0}$ for a field
dilution, in such a way that at each site the field $h_{i}$ obeys
$P(h_{i})=p_{+}\delta(h_{i}-h_{0})+p_{0}\delta(h_{i})
+p_{-}\delta(h_{i}+h_{0})$. The model is studied through the
replica method and phase diagrams are obtained within the
replica-symmetry approximation. It is shown that the border of
the ferromagnetic phase may present, for conveniently chosen values of
$p_{0}$ and $h_{0}$, first-order phase transitions, as well as tricritical
points at finite temperatures. Analogous to what happens for the Ising
ferromagnet under a trimodal random field,
it is verified that the first-order phase transitions are directly
related to the dilution in the fields: the extensions of these transitions
are reduced for increasing values of $p_{0}$.
Whenever the delta function at the origin becomes comparable to those
at $h_{i}=\pm h_{0}$, first-order phase transitions disappear; in fact, the
threshold value $p_{0}^{*}$, above which all phase transitions are continuous,
is calculated analytically as
$p_{0}^{*}=2(e^{3/2}+2)^{-1} \approx 0.30856$. The ferromagnetic boundary
at zero temperature also exhibits an interesting behavior: for
$0<p_{0}<p_{0}^{*}$, a single tricritical point occurs, whereas if
$p_{0}>p_{0}^{*}$ the critical frontier is completely continuous; however,
for $p_{0}=p_{0}^{*}$, a fourth-order critical point appears. The stability
analysis
of the replica-symmetric solution is performed and the regions of
validity of such a solution are identified; in particular, the
Almeida-Thouless line in
the plane field versus temperature is shown to depend on the weight $p_{0}$.

\vspace{2cm}


\end{abstract}

\newpage

{\large\bf 1. \quad Introduction}

\vskip \baselineskip

Among disordered magnets \cite{dotsenkoreview}, spin glasses
\cite{youngbook,fischerhertz,binderyoung} and ferromagnets in the presence
of random fields
\cite{nattermannvillain,nattermannrujan,belangeryoung,nattermann98}
may be singled out as two of the most puzzling and controversial systems
in condensed matter physics.

The random-field Ising model (RFIM), introduced by Imry and Ma \cite{imryma},
has concentrated a lot of interest after the identification of its physical
realizations. Probably the most important physical conception of the RFIM
comes out to be a diluted Ising antiferromagnet in the presence of a uniform
magnetic field \cite{fishmanaharony,cardy}. Since then, many diluted
antiferromagnets have been investigated, in such a way that systems like
${\rm Fe_{x}Zn_{1-x}F_{2}}$ and ${\rm Fe_{x}Mg_{1-x}Cl_{2}}$ are nowadays
considered as standard experimental realizations of the RFIM
\cite{belanger92,belanger98}. From the theoretical point of view, many
important ingredients remain unknown. At the mean-field level, it is well
known that different probability distributions for the random fields may
lead to distinct phase diagrams, e.g., a Gaussian probability
distribution yields a continuous ferromagnetic-paramagnetic boundary
\cite{schneiderpytte}, whereas for a bimodal distribution, this boundary
exhibits a continuous piece at high temperatures ending up at a tricritical
point, which is followed by a first-order phase transition at low
temperatures \cite{aharony78}. Such a contrast in the mean-field phase
diagrams of the RFIM with the bimodal and Gaussian probability distributions
has been proven rigorously \cite{salinas85}.
Indeed, Aharony \cite{aharony78} argued that
whenever an analytic symmetric distribution for the fields presents a minimum
at zero
field, one should expect a tricritical point and a first-order transition
for sufficiently low temperatures. Further studies of the RFIM at the
mean-field level have considered a trimodal (three-peak) distribution
\cite{mattis,kaufman86}

\vspace{-5mm}

$$
P(h_{i})=p_{+}\delta(h_{i}-h_{0})+p_{0}\delta(h_{i})
+p_{-}\delta(h_{i}+h_{0})
~, \eqno(1.1)
$$

\vspace{-5mm}

\vskip \baselineskip
\noindent
in its symmetrical form, i.e., $p_{+}=p_{-}={1 \over 2}(1-p_{0})$.
Such a distribution, which may be
interpreted as a bimodal added to a dilution in the fields with probability
$p_{0}$ \cite{mattis}, is expected to mimic better real systems
than its bimodal counterpart. It was shown that the field dilution
plays an important role in what concerns the presence
of the tricritical point: distinct analyses lead to slightly different
estimates for the threshold value, above which the tricritical point
disappears (whereas the analysis of Mattis \cite{mattis} shows that the
tricritical point vanishes for $p_{0}>0.25$, according to Kaufman
et al. \cite{kaufman86} such a behavior should occur for $p_{0}>0.24$).
Whether the features in the
mean-field phase diagrams of the RFIM should prevail on
short-range-interaction models,
represents a point which has attracted a lot of interest
\cite{riegeryoung93,rieger95,gofman,swift}. For the three-dimensional RFIM,
recent Monte Carlo simulations detect a jump in the magnetization but no
latent heat, for both bimodal \cite{riegeryoung93} and Gaussian
\cite{rieger95} distributions, whereas high-temperature series expansions
\cite{gofman} and a zero-temperature scaling analysis \cite{swift} find a
continuous transition for both distributions. However, in four dimensions
the same zero-temperature analysis \cite{swift} leads to a first-order
transition in the bimodal case and a continuous one for a Gaussian
distribution, in agreement with the mean-field predictions. Apart from that,
the low-temperature phase of the RFIM, in finite dimensions, may present a
nontrivial structure, with a complicated free-energy landscape, as suggested
by perturbative analyses \cite{mezardyoung,dominicis95}.

The Ising spin-glass (ISG) problem became, nowadays, one of the most
controversial
issues in the physics of disordered magnets. Its mean-field theory,
based on the solution of the infinite-range-interaction model,
the so-called Sherrington-Kirkpatrick (SK) model \cite{sk}, presents a quite
nontrivial behavior. The correct low-temperature solution, as proposed
by Parisi \cite{parisirsb}, consists of a continuous order-parameter function
(i.e., an infinite number of order parameters) associated with many
low-energy states, a procedure which is usually
denominated as replica-symmetry breaking (RSB). Furthermore, a transition
in the presence of an external magnetic field, known as the Almeida-Thouless
(AT) line \cite{at}, is found in the solution of the SK model: such a line
separates a low-temperature region, characterized by RSB, from a
high-temperature one, where a simple one-parameter solution, denominated as
replica-symmetric (RS) solution, is stable. The validity of the results of
the SK model for the description of real (short-range-interaction) systems
represents a very polemic question \cite{youngbook}. The rival theory is
the droplet model \cite{fisherhuse}, based on domain-wall
renormalization-group arguments for spin glasses
\cite{mcmillan84,braymoore87}. According to the droplet model, the
low-temperature phase of any \underline{finite-dimensional} short-range spin
glass should be described in terms of a single thermodynamic state
(together, of course, with its corresponding time-reversed counterpart), i.e.,
essentially a RS-type of solution. Obviously, the droplet model becomes
questionable for increasing dimensionalities, where one expects the existence
of a finite upper critical dimension -- believed to be six for the ISG
\cite{dominicis98} -- above which the mean-field picture should prevail.
Recent analyses of short-range ISG on diamond hierarchical lattices
(on which the Migdal-Kadanoff renormalization group is exact) has
found evidences of the droplet picture \cite{moore98}; however, the
applicability of such lattices for the description of ISG on Bravais lattices
is doubtful \cite{marinari98,marinari99}. Numerical simulations are very
hard to be carried for short-range ISG on a cubic lattice, due to large
thermalization times \cite{marinari98}; as a consequence, no conclusive
results in three-dimensional systems are available. However, in four
dimensions the critical temperature is much higher, making thermalization
easier; in this case, many works claim to have observed some mean-field
features \cite{mcdg3}.

From the theoretical point of view these two problems (RFIM and ISG), have
been, in most of the cases, studied in separate, with a few exceptions
\cite{pirc,ma,soares,nogueira98,vieira99}. However, many diluted
antiferromagnets, like
${\rm Fe_{x}Zn_{1-x}F_{2}}$ \cite{montenegro} and
${\rm Fe_{x}Mg_{1-x}Cl_{2}}$ \cite{bertrand,wong}, are able to
exhibit, within certain concentration ranges, random-field, spin-glass or
both behaviors. For the
${\rm Fe_{x}Zn_{1-x}F_{2}}$, one gets a RFIM for ${\rm x} \ge 0.40$, an ISG
for ${\rm x} \le 0.24$, whereas for intermediate concentrations
($0.24 \le {\rm x} \le 0.40$) one may observe both behaviors depending on
the magnitude of the applied external magnetic field [RFIM (ISG) for small
(large) magnetic fields], with a crossover between them; this latter effect
was observed in ${\rm Fe_{0.31}Zn_{0.69}F_{2}}$ \cite{montenegro}. Certainly,
such properties are expected to be properly explained only if one considers a model
which takes into account both spin-glass and random-field ingredients.
Indeed, the crossover observed in ${\rm Fe_{0.31}Zn_{0.69}F_{2}}$ was also
found in the study of the SK model under a Gaussian random field
\cite{soares}. On the other hand, the study of the SK model in the presence
of a bimodal random field produced interesting results,
with first-order phase transitions and tricritical points \cite{nogueira98};
such results may be relevant for explaining the first-order phase transitions
observed in ${\rm Fe_{x}Mg_{1-x}Cl_{2}}$ \cite{belanger98}.

In the present work we study the SK model in the presence of a random field
following a trimodal probability distribution [see Eq. (1.1)].
In addition to that, one may
interpolate between the bimodal distribution and a behavior which is
qualitatively analogous to the Gaussian one, since by monitoring the delta
function at the origin, one is able to control the presence of tricritical
points. In the next section we define the model and, through the use of the
replica method, we find its free-energy density, equations of state and
equations for the validity of the RS solution. In
section 3 we exhibit and discuss the phase diagrams of the model. Finally,
in section 4 we present our conclusions.

\vskip 2\baselineskip

{\large\bf 2. \quad The Model and Replica Formalism}

\vskip \baselineskip

The mean-field theory of the ISG is usually formulated as a set
of $N$ spins, each of them interacting with all others [a total
of ${1 \over 2}N(N-1)$ interactions], known as the SK model \cite{sk}.
The SK model
in the presence of an external random magnetic field may be
defined in terms of the Hamiltonian \cite{soares,nogueira98},

$$
{\cal H} = -\sum_{(ij)} J_{ij} S_{i}S_{j} - \sum_{i} h_{i}S_{i}
~, \eqno(2.1)
$$

\vskip \baselineskip
\noindent
where $S_{i}=\pm 1 \ $, with $i=1,2, \cdots ,N$, and the interactions are
infinite-range-like, i.e., the sum $\sum_{(i,j)}$
applies to all distinct pairs of spins. The coupling constants $\{J_{ij}\}$ and the
random fields $\{h_{i}\}$ are quenched variables, following independent
probability distributions,

$$
P(J_{ij}) = \left({N \over 2\pi J^2}\right)^{1\over2} \exp \left[-{N
\over 2 J^2} \left( J_{ij} - {J_0 \over N} \right)^2 \right]
~, \eqno(2.2)
$$

\vskip \baselineskip
\noindent
with $P(h_{i})$ given by Eq. (1.1) \ ($p_{+}+p_{0}+p_{-}=1$). Let us, for
the moment, keep the trimodal probability distribution in its general form
of Eq. (1.1); later on, we will see that the ferromagnetic boundary does
not exist
for $p_{+} \ne p_{-}$, and so, in such a case, we will be restricted to the
symmetrical form
$p_{+}=p_{-}={1 \over 2}(1-p_{0})$. It should be mentioned that the above
randomnesses ($\{J_{ij}\}$ and $\{h_{i}\}$) are usually correlated in real
systems; herein for the sake of simplicity, we shall consider two independent
probability distributions.
Therefore, for a given realization of bonds and site-fields,
$(\{J_{ij}\},\{h_{i}\})$, one
has a corresponding free energy, $F(\{J_{ij}\},\{h_{i}\})$, such that the
average over the disorder, $[ \ \ ]_{J,h}$, may be performed as independent
integrals,

$$
[F(\{J_{ij}\},\{h_{i}\})]_{J,h} = \int \prod_{(ij)} [dJ_{ij}P(J_{ij})]
\ \prod_{i} [dh_{i}P(h_{i})] \ F(\{J_{ij}\},\{h_{i}\})
~. \eqno(2.3)
$$

\vskip \baselineskip
The usual procedure consists in applying the replica method
\cite{fischerhertz,binderyoung}, in such a way as to get the free energy per
spin as,

\setcounter{enumi}{2}
\setcounter{equation}{3}
\renewcommand{\theequation}{\arabic{enumi}.\arabic{equation}}
\begin{eqnarray}
- \beta f & = & \lim_{N \rightarrow \infty} {1 \over N} \
[ \, \ln Z(\{J_{ij}\},\{h_{i}\})]_{J,h} \nonumber \\
& = & \lim_{N \rightarrow \infty} \lim_{n \rightarrow 0} {1 \over Nn}
\left( [Z^n]_{J,h} - 1 \right)
~,
\end{eqnarray}

\vskip \baselineskip
\noindent
where $Z^n$ is the partition function of $n$ copies of the system defined
in Eq. (2.1) and $\beta = 1/T$ \ (we work in units $k_{B}=1$). Standard
calculations lead to

$$
\beta f = - {(\beta J)^{2} \over 4} +
\lim_{n \rightarrow 0} \ {1\over n} \min \, g(m^\alpha ,q^{\alpha \beta })
~, \eqno(2.5)
$$

\vskip \baselineskip
\noindent
where

$$
g(m^\alpha ,q^{\alpha \beta }) = {\beta J_{0} \over 2} \sum _{\alpha}
 \ (m^\alpha )^{2} + {(\beta J)^{2} \over 2}
\sum _{(\alpha \beta)} \left(q^{\alpha \beta }\right)^{2}
- \ p_{+} \ln \ {\rm Tr}_{\alpha} \exp ({\cal H}_{eff}^{+})
\nonumber
$$

$$
- \ p_{0} \ln \ {\rm Tr}_{\alpha} \exp ({\cal H}_{eff}^{0})
- \ p_{-} \ln \ {\rm Tr}_{\alpha} \exp ({\cal H}_{eff}^{-})
~, \eqno(2.6 {\rm a})
$$

$$
{\cal H}_{eff}^{\pm} = \beta J_{0} \sum _{\alpha} m^{\alpha} \ S^{\alpha} +
(\beta J)^2 \sum _{(\alpha \beta)} q^{\alpha \beta } \ S^{\alpha}S^{\beta}
\pm \ \beta h_{0} \sum _{\alpha} S^{\alpha}
~, \eqno(2.6 {\rm b})
$$

$$
{\cal H}_{eff}^{0} = \beta J_{0} \sum _{\alpha} m^{\alpha} \ S^{\alpha} +
(\beta J)^2 \sum _{(\alpha \beta)} q^{\alpha \beta } \ S^{\alpha}S^{\beta}
~. \eqno(2.6 {\rm c})
$$

\vskip \baselineskip
\noindent
In the equations above, the sum indexes
$\alpha$ and $\beta$ $(\alpha, \beta = 1,2,\cdots,n)$
are replica labels and $\sum_{(\alpha \beta)}$ denote sums over distinct pairs
of replicas.

The extrema of the functional $g(m^\alpha ,q^{\alpha \beta })$ give us the
equilibrium equations for the magnetization and spin-glass order parameters,
respectively,

\setcounter{enumi}{2}
\setcounter{enumii}{7}
\setcounter{equation}{0}
\renewcommand{\theequation}{\arabic{enumi}.\arabic{enumii}\alph{equation}}
\begin{eqnarray}
m^\alpha & = & p_{+} \langle S^{\alpha } \rangle _{+}
+ p_{0} \langle S^{\alpha } \rangle _{0}
+ p_{-} \langle S^{\alpha } \rangle _{-}
~, \\
q^{\alpha \beta } & = & p_{+} \langle S^{\alpha}S^{\beta}\rangle _{+}
+p_{0} \langle S^{\alpha}S^{\beta}\rangle _{0}
+p_{-} \langle S^{\alpha}S^{\beta}\rangle _{-}
\quad
\quad (\alpha \neq \beta) ~,
\end{eqnarray}

\noindent
where $\langle \; \rangle _{\pm}$ and $\langle \; \rangle _{0}$ refer to
thermal averages with respect to the ``effective Hamiltonians''
${\cal H}_{eff}^{\pm}$ and ${\cal H}_{eff}^{0}$ in Eqs. (2.6b) and (2.6c),
respectively.

If one assumes the replica-symmetry (RS) ansatz \cite{sk},

\vspace{-5mm}
$$
m^{\alpha} = m~, \quad \forall \alpha \quad; \quad
q^{\alpha \beta} = q~, \quad \forall (\alpha \beta)
~, \eqno(2.8)
$$

\noindent
the free energy per spin (Eq. (2.5)) and the equilibrium conditions
(Eqs. (2.7)) become

$$
\beta f = - {(\beta J)^{2} \over 4} (1 - q)^{2}
+ {\beta J_{0} \over 2} \ m^2
- \ p_{+} \int {\cal D} z \ln(2 \cosh \xi ^{+}) \
\nonumber
$$

$$
- \ p_{0} \int {\cal D} z \ln(2 \cosh \xi ^{0})
- \ p_{-} \int {\cal D} z \ln(2 \cosh \xi ^{-})
~, \eqno(2.9)
$$

$$
m = \ p_{+} \int {\cal D} z \tanh \xi ^{+} \
+ \ p_{0} \int {\cal D} z \tanh \xi ^{0}
+ \ p_{-} \int {\cal D} z \tanh \xi ^{-}
~, \eqno(2.10)
$$

$$
q = \ p_{+} \int {\cal D} z \tanh^{2} \xi ^{+} \
+ \ p_{0} \int {\cal D} z \tanh^{2} \xi ^{0}
+ \ p_{-} \int {\cal D} z \tanh^{2} \xi ^{-}
~, \eqno(2.11)
$$

\vskip \baselineskip
\noindent
where

$$
\int {\cal D} z \cdots = \int_{-\infty}^{\infty}
\left( {1 \over 2\pi} \right)^{1 \over 2} dz \ \exp (- z^{2}/2)
\ \cdots
\qquad , \eqno(2.12)
$$

\vskip \baselineskip
\noindent
and

\vspace{-5mm}
\setcounter{enumi}{2}
\setcounter{enumii}{13}
\setcounter{equation}{0}
\renewcommand{\theequation}{\arabic{enumi}.\arabic{enumii}\alph{equation}}
\begin{eqnarray}
\xi ^{\pm} & = & \beta J_{0} m + \beta J q ^{1/2} z \pm \beta h_{0}
~, \\
\xi ^{0} & = & \beta J_{0} m + \beta J q ^{1/2} z
~.
\end{eqnarray}

\vskip \baselineskip

Although the spin-glass order parameter (Eq. (2.11)) is always induced by
a nonzero random field ($p_{0}<1$), it may still contribute to a
nontrivial behavior; this is provided by the instability of the RS solution.
Such an instability occurs at the AT line \cite{at},

$$
\left( {T \over J} \right)^{2} =
\ p_{+} \int {\cal D} z \ {\rm sech}^{4} \xi ^{+} \
+ \ p_{0} \int {\cal D} z \ {\rm sech}^{4} \xi ^{0}
+ \ p_{-} \int {\cal D} z \ {\rm sech}^{4} \xi ^{-}
~, \eqno(2.14)
$$

\vskip \baselineskip
\noindent
which may be obtained through the simultaneous solution of Eqs. (2.14), (2.10)
and (2.11).

In the next section we shall consider the phase diagrams of the model and
the regions of instability of the RS solution, worked out from Eqs.
(2.9)--(2.14).

\vskip 2\baselineskip

{\large\bf 3. \quad Results and Discussion}

\vskip \baselineskip

Let us first consider the case $J_{0}=0$; one may easily see that the only
nontrivial behavior in this case is given by the AT instability in the plane
magnetic field versus temperature, which may
now be obtained from the solution of Eqs. (2.11) and (2.14).
The integrals involving $\xi ^{-}$ may
be easily transformed through the change of variables
$z \rightarrow -z \ $, in such a way that the AT line may be obtained by
solving the set of equations,

\setcounter{enumi}{3}
\setcounter{enumii}{1}
\setcounter{equation}{0}
\renewcommand{\theequation}{\arabic{enumi}.\arabic{enumii}\alph{equation}}
\begin{eqnarray}
\left( {T \over J} \right)^{2} & = & \ (1-p_{0}) \ \int {\cal D} z \
{\rm sech}^{4} (\beta J q ^{1/2} z + \beta h_{0})
+ \ p_{0} \ \int {\cal D} z \
{\rm sech}^{4} (\beta J q ^{1/2} z)
~, \\
\nonumber \\
q & = & \ (1-p_{0}) \ \int {\cal D} z \
{\rm tanh}^{2} (\beta J q ^{1/2} z + \beta h_{0})
+ \ p_{0} \ \int {\cal D} z \
{\rm tanh}^{2} (\beta J q ^{1/2} z)
~.
\end{eqnarray}

\noindent
It should be pointed out that the equations above are valid for arbitrary
values of the weights in the probability distribution of Eq. (1.1), with
$p_{+}+p_{-}=1-p_{0} \ $; although the AT line changes with field dilution,
it is no altered under a field inversion. The AT lines in the plane magnetic
field versus
temperature are exhibited in Fig. 1, for typical values of $p_{0}$.
Clearly, the AT line for the bimodal distribution ($p_{0}=0$)
\cite{nogueira98}
is identical to the one of the SK model in the presence of a uniform magnetic
field \cite{at}, due to the property of invariance under field inversion.
For $0<p_{0}<1$, one may calculate analytically the
behavior of the AT line in the low-field regime ($T \cong J$),

$$
1-{T \over J} \cong \left[ {3(1-p_0) \over 4} \right]^{1/3}
\left( {h_{0} \over J} \right)^{2/3}
~, \eqno(3.2)
$$

\vskip \baselineskip
\noindent
which leads to a slightly modified amplitude, but the same low-field
exponent of the standard AT line \cite{at}. If one considers
$p_{0} \sim 0$, the low-temperature behavior of the AT line may be easily
calculated,

$$
{T \over J} \cong {4 \over 3}{1 \over \sqrt{2 \pi}}
 \left[ (1-p_0) {\rm exp} \left( -{h_{0}^{2} \over 2J^{2}} \right)
+ p_{0} \right]
~, \eqno(3.3)
$$

\vskip \baselineskip
\noindent
which exhibits the usual exponential decay \cite{at}, but with a shift
towards higher temperatures for increasing values of $p_{0}$. In all other
situations, the AT lines were calculated by solving numerically Eqs. (3.1).
One notices that for high values
of $p_{0}$, the integrals multiplying $p_{0}$ in Eqs. (3.1) contribute
significantly, in such a way that the AT lines become slightly independent
of $h_{0}$, for $h_{0}$ large enough, as shown in Fig. 1.

From now on, we will be restricted to $J_{0}>0$; in this case, as far as
RS is concerned, if
$p_{+} \ne p_{-}$ Eqs. (2.10) and (2.11) yield nonzero magnetization and
spin-glass order parameters, leading to trivial behavior.
Therefore, for the rest of
this paper we will concentrate on a symmetrical trimodal distribution, i.e.,
$p_{+}=p_{-}={1 \over 2}(1-p_{0})$.
In this case, the random field still induces
the parameter $q$, leading to no spontaneous spin-glass order
(like the one found for the SK model in the absence of external field
\cite{sk}). Therefore, the only possible phase
transition within the RS approximation is the one associated with the
magnetization, similarly to what happened in the case of the bimodal
distribution \cite{nogueira98}. Hence, two phases are possible, namely,
the ferromagnetic ($m \ne 0 \ , \ q \ne 0$) and the independent
($m = 0 \ , \ q \ne 0$) ones. Although in the RFIM this latter phase is
usually denominated of paramagnetic, in the present problem, within the RS
approximation, we shall keep the nomenclature independent, for reasons
which will become clear soon.

The critical frontier separating these two phases may be found by solving
the equilibrium equations, (2.10) and (2.11); in the case of first-order
phase transitions, we shall make use of the free-energy per spin
[Eq. (2.9)] as well. Expanding Eq. (2.10) in powers of $m$ one gets,

\vspace{-5mm}

$$
m = A_{1}(q) \ m + A_{3}(q) \ m^{3} + A_{5}(q) \ m^{5} + O(m^{7})
~, \eqno(3.4)
$$

\noindent
where the coefficients depend on $q$ [which on its turn, depends on $m$
through Eq. (2.11)]. Expanding Eq. (2.11) in powers of $m$,

$$
q = q_{0} + \ {(\beta J_{0})^{2} \Gamma \over
1 - (\beta J)^{2}  \ \Gamma} \ m^{2} + O(m^{4})
~, \eqno(3.5)
$$

\vskip \baselineskip
\noindent
with

$$
\Gamma = (1-p_{0})(1-4\rho_{1}^{+}+3\rho_{2}^{+})
+p_{0}(1-4\rho_{1}^{0}+3\rho_{2}^{0})
~, \eqno(3.6)
$$

\setcounter{enumi}{3}
\setcounter{enumii}{7}
\setcounter{equation}{0}
\renewcommand{\theequation}{\arabic{enumi}.\arabic{enumii}\alph{equation}}
\begin{eqnarray}
\rho_{k}^{+} & = & \int {\cal D} z \ {\rm tanh}^{2k} (\beta J q_{0}^{1/2} z
+ \beta h_{0})
~, \\
\rho_{k}^{0} & = & \int {\cal D} z \ {\rm tanh}^{2k} (\beta J q_{0}^{1/2} z)
~,
\end{eqnarray}

\vspace{-5mm}

\vskip \baselineskip
\noindent
where $q_{0}$ is independent of $m$, corresponding to the solution of
Eq. (2.11) with $m=0$. Substituting the above results into Eq. (3.4),
one gets the $m$-independent coefficients of the power expansion,

\setcounter{enumi}{3}
\setcounter{enumii}{8}
\setcounter{equation}{0}
\renewcommand{\theequation}{\arabic{enumi}.\arabic{enumii}\alph{equation}}
\begin{eqnarray}
A_{1}^{\prime} & = & \beta J_{0} [1-(1-p_{0})\rho_{1}^{+}-p_{0}\rho_{1}^{0}]
~, \\ \nonumber \\
A_{3}^{\prime} & = & - {(\beta J_{0})^{3} \over 3} \ \left[
{1 + 2(\beta J)^{2}  \ \Gamma \over 1 - (\beta J)^{2}  \ \Gamma} \right]
\ \Gamma
~, \\ \nonumber \\
A_{5}^{\prime} & = & -\gamma \ {(\beta J_0)^5\over 30} \left [
{1+8(\beta J)^2 \ \Gamma + 36(\beta J)^4 \ \Gamma^2
+ 15(\beta J)^6 \ \Gamma^3 \over
1-(\beta J)^2 \ \Gamma } \right ]
~,
\end{eqnarray}

\vspace{-5mm}

\vskip \baselineskip
\noindent
where

\vspace{-5mm}

$$
\gamma = (1-p_{0})(-4+34\rho_{1}^{+}-60\rho_{2}^{+}+30\rho_{3}^{+})
+p_{0}(-4+34\rho_{1}^{0}-60\rho_{2}^{0}+30\rho_{3}^{0})
~. \eqno(3.9)
$$

\vspace{-5mm}

\vskip \baselineskip
\noindent
The critical frontier may be determined using standard procedures,
as described below.

(i) For continuous phase transitions, $A_{1}^{\prime}=1$
and $A_{3}^{\prime} <0 \ $.

(ii) A first-order phase transition occurs whenever $A_{1}^{\prime}=1$ and
$A_{3}^{\prime}>0$; the proper critical frontier should be found, in this
case, through a
Maxwell construction, i.e., by equating the free energies of the two phases.

(iii) When both types of phase transitions are present, the continuous- and
first-order critical frontiers meet at a tricritical point
\cite{lawriesarbach}, which defines
the limit of validity of the series expansions; beyond the tricritical
point the magnetization is discontinuous. The location of such
point is determined by setting
$A_{1}^{\prime}=A_{3}^{\prime}=0$, with the condition $A_{5}^{\prime}<0$
satisfied.

In Figs. 2--4 we show three qualitatively distinct ferromagnetic boundaries
of the present problem, for a typical value of $p_{0}$ \ ($p_{0}=0.3$),
compared with those of the bimodal probability
distribution ($p_{0}=0$). In Fig. 2 there is a single point along the
ferromagnetic boundary at which $A_{3}^{\prime}=0$; such a point may not be
considered as tricritical, since there is no first-order phase
transition. However, for any value of $h_{0}$
greater than those of Fig. 2
[$h_{0}/J=0.9573$ \ ($p_{0}=0$) and $h_{0}/J=1.53526$ \ ($p_{0}=0.3$)], one
gets first-order phase transitions, and at least one tricritical point.
In Fig. 3 we show situations where two tricritical points appear along the
ferromagnetic boundary; we have
verified that, for a fixed value of $p_{0}$, such a behavior occurs within
a narrow interval of $h_{0}$. In Fig. 4 a single
tricritical point emerges, separating a continuous boundary (high
temperatures) from a first-order critical frontier (low temperatures). From
such phase diagrams, one notices that the main effect of the field dilution
is to push the tricritical points
towards lower temperatures, i.e.,
the temperature range over which the first-order transitions occur decreases.

As mentioned before, although the spin-glass order parameter is always
induced by the random field, it may still exhibit interesting behavior,
associated with the instability of the RS solution. The AT instabilities,
given by the solution of Eqs. (2.10), (2.11) and (2.14) with
$p_{+}=p_{-}={1\over2}(1-p_{0})$, yields two distinct lines in the phase
diagrams of Figs. 2--4, depending on whether one is inside the independent
phase ($m=0$), or in the ferromagnetic ($m \ne 0$) one. In the former case,
the AT line is a straight line (independent of $J_{0}$), whereas in the
latter, it presents the usual decrease with temperature for increasing values
of $J_{0}$, in such a way that for low temperatures one gets the exponential
decays,

$$
{T \over J} \cong {4 \over 3} {1 \over \sqrt{2 \pi}}
\left \{
{1\over2}(1-p_{0}) \ \exp \left [ - {(J_{0} + h_{0})^{2} \over 2J^{2}} \right]
+p_{0} \ \exp \left [ - {J_{0}^{2} \over 2J^{2}} \right] \right.
\nonumber
$$

$$
\left.
+{1\over2}(1-p_{0}) \ \exp \left [ - {(J_{0} - h_{0})^{2} \over 2J^{2}} \right]
\right \}
~. \eqno(3.10)
$$

\vskip \baselineskip
\noindent
Herein we shall adopt the usual criteria for the identification of the
regions where RS is stable and those throughout which a RSB procedure
is necessary \cite{fischerhertz,binderyoung}. The two regions with
zero magnetization
will be associated with the paramagnetic (high temperatures) and spin-glass
(low temperatures) phases, whereas those with nonzero magnetization will be
associated with the ferromagnetic (high temperatures) and mixed-ferromagnetic
(low temperatures). The several phases exhibited in our phase diagrams are
identified as:

\vskip \baselineskip

\begin{tabular}{lll}
Paramagnetic ({\bf P}) & ($m=0$ \ ; \ $q$ \ : \ RS) & ; \\
Spin-Glass ({\bf SG}) & ($m=0$ \ ; \ $q$ \ : \ RSB) & ; \\
Ferromagnetic ({\bf F}) & ($m \ne 0$ \ ; \ $q$ \ : \ RS) & ; \\
Mixed Ferromagnetic ({\bf F}$^{\prime}$) & ($m \ne 0$ \ ; \ $q$ \ : \ RSB) & .
\end{tabular}

\vskip \baselineskip

It should be mentioned that the present low-temperature results are
questionable inside the
phases {\bf F}$^{\prime}$ and {\bf SG}, due to the instability
of the RS solution; in particular the point for $p_{0}=0.3$ where
$A_{3}^{\prime}=0$ in Fig. 2, as well as the low-temperature tricritical
points of Fig. 3 may completely disappear under a RSB procedure. However
the high-temperature tricritical points, like those of Figs. 3 and 4, are
inside the region of stability of the RS solution and will persist
under more general treatments; we believe that such points are reminiscent
of the tricritical point of the bimodal RFIM.

The two AT lines mentioned above usually meet at a continuous ferromagnetic
boundary; however, these
lines do not match each other across first-order phase transitions
\cite{nogueira98,vieira99,fyodorov87a}: there is a small (but finite)
gap between them in Figs. 3 and 4.

Let us now investigate the ferromagnetic boundary at zero
temperature; for $T=0$ the spin-glass order parameter is trivial ($q=1$),
in such a way that one gets for the free energy and magnetization,

\setcounter{enumi}{3}
\setcounter{enumii}{11}
\setcounter{equation}{0}
\renewcommand{\theequation}{\arabic{enumi}.\arabic{enumii}\alph{equation}}
\begin{eqnarray}
f & = & - {J_{0} \over 2} m^{2} - {h_{0} \over 2} (1-p_{0}) \left[
{\rm erf} \left( {J_{0}m + h_{0} \over J \sqrt{2}} \right)
- {\rm erf} \left( {J_{0}m - h_{0} \over J \sqrt{2}} \right) \right]
\nonumber \\ \nonumber \\
& & - {J \over \sqrt{2 \pi}} (1-p_{0}) \left\{
\exp \left[ - {(J_{0}m + h_{0})^{2} \over 2J^{2}} \right]
+ \exp \left[ - {(J_{0}m - h_{0})^{2} \over 2J^{2}} \right] \right\}
\nonumber \\ \nonumber \\
& & - {2J \over \sqrt{2 \pi}} \ p_{0} \left\{
\exp \left[ - {(J_{0}m)^{2} \over 2J^{2}} \right] \right\}
~, \\
\nonumber \\
m & = & {1 \over 2} (1-p_{0}) \left[
{\rm erf} \left( {J_{0}m + h_{0} \over J \sqrt{2}} \right)
+ {\rm erf} \left( {J_{0}m - h_{0} \over J \sqrt{2}} \right) \right]
+ p_{0} \ {\rm erf} \left( {J_{0}m \over J \sqrt{2}} \right)
~.
\end{eqnarray}

\vspace{-5mm}

\vskip \baselineskip
\noindent
Using a similar procedure as the one for finite temperatures, one may expand
Eq. (3.11b),

\vspace{-5mm}

$$
m = a_{1} \ m + a_{3} \ m^{3} + a_{5} \ m^{5} + O(m^{7})
~, \eqno(3.12)
$$

\noindent
where,

\setcounter{enumi}{3}
\setcounter{enumii}{13}
\setcounter{equation}{0}
\renewcommand{\theequation}{\arabic{enumi}.\arabic{enumii}\alph{equation}}
\begin{eqnarray}
a_{1} & = & \sqrt{{2 \over \pi}} \ {J_{0} \over J} \ \left[
(1-p_{0}) \exp \left( - {h_{0}^{2} \over 2J^{2}} \right) + p_{0} \right]
~, \\
\nonumber \\
a_{3} & = & {1 \over 6} \ \sqrt{{2 \over \pi}} \
\left( {J_{0} \over J} \right)^{3} \left[
(1-p_{0}) \left( {h_{0}^{2} \over J^{2}} - 1 \right)
\exp \left( - {h_{0}^{2} \over 2J^{2}} \right) - p_{0} \right]
~, \\
\nonumber \\
a_{5} & = & {1 \over 120} \ \sqrt{{2 \over \pi}} \
\left( {J_{0} \over J} \right)^{5} \left[
(1-p_{0}) \left( {h_{0}^{4} \over J^{4}} - 6 \ {h_{0}^{2} \over J^{2}} + 3
\right)
\exp \left( - {h_{0}^{2} \over 2J^{2}} \right) - 3p_{0} \right]
~.
\end{eqnarray}

\vskip \baselineskip
\noindent
The critical frontier separating the phases {\bf F}$^{\prime}$ and {\bf SG}
is shown in Fig. 5 for typical values of $p_{0}$.
One notices that the effect of the weight $p_{0}$ is to favour the continuous
line, along which $a_{1}=1$ with $a_{3}<0$, i.e.,

$$
{J_{0} \over J} = \sqrt{{\pi \over 2}} \ {1 \over
p_{0} + (1-p_{0}) \exp (-h_{0}^{2} / 2J^{2})}
~, \eqno(3.14)
$$

\vskip \baselineskip
\noindent
while decreasing the extension of the first-order transition line. For small
values of $p_{0}$ these two lines meet at a tricritical point, obtained by
solving the equations $a_{1}=1$, $a_{3}=0$, with the condition $a_{5}<0$;
within the analysis for finite temperatures,
this corresponds to the situation where the lower-temperature tricritical
point (cf. Fig. 3) hits the zero-temperature axis.
If $p_{0}=0$ such an effect occurs at \cite{nogueira98}

$$
{h_{0} \over J} = 1 \quad ; \quad {J_{0} \over J} =
\sqrt{{\pi e \over 2}} \approx 2.0664
~. \eqno(3.15)
$$

\vskip \baselineskip
\noindent
We verified that for $0<p_{0}<p_{0}^{*}$ \ (where $p_{0}^{*}$ will be defined
below), such a set of equations presents two solutions, although only
one of them represents a tricritical point, satisfying $a_{5}<0$. By
increasing $p_{0}$ inside this range, we noticed that such solutions
get closer and colapse for $p_{0}=p_{0}^{*}$. We calculated analytically
$p_{0}^{*}=2(e^{3/2}+2)^{-1} \approx 0.30856$, at which a fourth-order
critical
point \cite{griffiths75} (characterized by
$a_{1}=a_{3}=a_{5}=0$, with $a_{7}<0$) occurs at

$$
{h_{0} \over J} = \sqrt{3} \approx 1.73207 \quad ; \quad
{J_{0} \over J} = { \sqrt{2 \pi} \over 6} (e^{3/2}+2) \approx 2.70786
~. \eqno(3.16)
$$

\vskip \baselineskip
\noindent
The value $p_{0}^{*}$ represents a threshold of $p_{0}$, above which
there are no first-order transitions for any temperature $T \ge 0$. For
$p_{0}>p_{0}^{*}$ the second-order critical frontier of Fig. 5 approaches
an asymptote for large values of $h_{0}$; indeed, when
$p_{0} \rightarrow 1$ the zero-temperature ferromagnetic boundary approaches
a straight line at $J_{0}/J=\sqrt{\pi /2}$ \ [see Eq. (3.14)], characteristic
of the SK model in zero field \cite{sk}.

It should be mentioned that the finite-temperature vestigial points where
$A_{3}^{\prime}=0$, like the ones in Fig. 2, are qualitatively different
from the fourth-order critical point found for $p_{0}=p_{0}^{*}$ at zero
temperature, even though both situations represent thresholds for the
occurrence of tricritical points. In the former case, $A_{5}^{\prime}<0$,
whereas in the latter, $A_{5}^{\prime}=0$. In Fig. 6 we exhibit the behavior
of the coefficients $A_{3}^{\prime}$ and $A_{5}^{\prime}$, for temperatures
along the ferromagnetic frontier, for the case (b) of Fig. 2, i.e.,
$p_{0}=0.3$ \ ($h_{0}/J=1.53526$), and $p_{0}=p_{0}^{*}$ \
($h_{0}/J=\sqrt{3}$). One clearly sees that the fourth-order critical point
only shows up at zero temperature; its parameters, as defined in Eq. (3.16),
correspond to the situation where the vestigial point of Fig. 2 collapses
with the zero-temperature axis.

If $0<p_{0}<p_{0}^{*}$, it is always possible to obtain first-order phase
transitions by conveniently choosing the value of $h_{0}$.
In Fig. 7 we
exhibit the ranges of $p_{0}$ and $h_{0}/J$ throughout which first-order
phase transitions and tricritical points are possible along the ferromagnetic
boundary. In region (a), first-order phase transitions are conceivable at
finite and zero temperature, with a single tricritical point (at finite
temperatures): typical examples are shown in Fig. 4. Throughout a very narrow
range [region (b)] two tricritical points appear and the first-order phase
transition occurs only for finite temperatures: typical examples are
exhibited in Fig. 3. The region (b) is delimited by characteristic values
of ($p_{0},h_{0}/J$):
(i) the threshold for $h_{0}/J$ smaller corresponds to the set of points
satisfying $A_{3}^{\prime}=0$, but with no first-order phase transition
(e.g., the vestigial points shown in Fig. 2);
(ii) the delimiter for $h_{0}/J$ larger corresponds to the coordinates of
the tricritical points at zero temperature.
The vertical line in Fig. 7 is for $p_{0}=p_{0}^{*}$, defining [together
with the delimiter (i) of region (b)], the range throughout which the
ferromagnetic boundary is always continuous [region (c)].

\vskip 2\baselineskip

{\large\bf 4. \quad Conclusion}

\vskip \baselineskip

We have studied the Sherrington-Kirkpatrick spin glass in the presence of
random fields $\{ h_{i} \} $, following a trimodal (three-peak) probability
distribution, which corresponds to a bimodal plus a probability $p_{0}$ for
field dilution, i.e.,
$P(h_{i})=p_{+}\delta(h_{i}-h_{0})+p_{0}\delta(h_{i})
+p_{-}\delta(h_{i}+h_{0})$.
We have used
the replica method and the phase diagrams were obtained within the
replica-symmetry approximation. The boundary of the ferromagnetic phase
exhibited an interesting behavior, with the presence of first-order phase
transitions and tricritical points: within certain ranges for $p_{0}$ and
$h_{0}$, a single or two tricritical points were encountered.
We have shown that the first-order
phase transitions are directly affected by the dilution in the fields,
in such a away that the extension
of such lines are reduced by increasing $p_{0}$. In fact, there is a
threshold value, $p_{0}^{*}=2(e^{3/2}+2)^{-1} \approx 0.30856$, above which
the ferromagnetic boundary is always continuous. Such effects may be
reminiscent of those occurring within the mean-field theory of the Ising
ferromagnet
in the presence of trimodal random fields: the single tricritical point
that appears
in the case of a bimodal distribution \cite{aharony78} is washed way by the
presence of the delta at the origin, whenever $p_{0}$ becomes greater than a
certain value \cite{mattis,kaufman86}.

At zero temperature, if $0<p_{0}<p_{0}^{*}$, the ferromagnetic
critical frontier exhibits a single tricritical point, with a first-order
phase transition at high values of $h_{0}$. By increasing $p_{0}$, the
first-order line gets reduced and, for $p_{0}=p_{0}^{*}$, a fourth-order
critical point is observed; for $p_{0}>p_{0}^{*}$, the ferromagnetic boundary
is always continuous.

Although the spin-glass order parameter is induced by the random
field ($p_{0}<1$), it may still contribute to a nontrivial behavior, in what
concerns the stability of the replica-symmetric solution. We have
calculated the regions of instability of such a solution,
leading to the identification of two low-temperature phases, namely, the
spin-glass and mixed ferromagnetic ones. Besides that, the Almeida-Thouless
line in the
plane field versus temperature was shown to depend on the weight
$p_{0}$, with different amplitudes (but the same exponent) in the low-field
regime, and qualitatively distinct high-field behaviors.

We have verified that whenever the ferromagnetic boundary presents both
continuous and first-order transition lines meeting at a single
finite-temperature tricritical point, such a point is
located inside the region of stability of the replica-symmetric solution,
and it will not be removed by a replica-symmetry-breaking
procedure. However, when two tricritical points occur along the ferromagnetic
boundary, at least one of them (the one at low temperatures) appears inside
the unstable region, and its existence may be an artifact of the
replica-symmetric solution.

The applicability of the present results in the description of real systems
obviously depends on the survival of the mean-field characteristics in the
respective short-range-interaction versions of Ising spin glasses and the
Ising ferromagnet in the presence of a random field. However, the
trimodal distribution employed herein is expected to mimic better real
systems than
the bimodal distribution itself. Although we are not aware of
experimental observations that match with our results, we believe that the
diluted antiferromagnet ${\rm Fe_{x}Mg_{1-x}Cl_{2}}$ is a good candidate,
since, for conveniently chosen dilutions, it may exhibit first-order
phase transitions \cite{belanger98}, as well as a crossover from first- to
second-order behavior \cite{kushhauerkleemann}.

\vskip 2\baselineskip
{\large\bf Acknowledgments}

\noindent
We acknowledge E.~M.~F. Curado for useful discussions. FDN thanks CNPq and
Pronex/MCT (Brazilian granting agencies) for partial financial support.

\newpage
\centerline{{\large\bf Figure Captions}}

\vskip 2\baselineskip
\noindent
{\bf Fig. 1:} \quad The AT lines, for the SK model in the presence of a
trimodal random field, in the plane $h_{0}$ versus $T$ (in units of $J$),
for typical values of $p_{0}$.

\vskip 2\baselineskip
\noindent
{\bf Fig. 2:} \qquad Phase diagram $T$ versus $J_{0}$ (in units of $J$) of
the SK model in the presence of a trimodal random field with $p_{0}=0.3$,
compared with one of the bimodal case ($p_{0}=0$), for conveniently
chosen values of $h_{0}$.
(a) $h_{0}/J=0.9573$ \ ($p_{0}=0$);
(b) $h_{0}/J=1.53526$ \ ($p_{0}=0.3$).
The ferromagnetic boundaries are continuous, except for the points where
$A_{3}^{\prime}=0$ \ [cf. Eq. (3.7b)], represented by black squares. These
choices signal lower bounds for $h_{0}$, above which first-order phase
transitions occur. The phase nomenclature is specified in the text, with
the low-temperature phases {\bf SG} and {\bf F}$^{\prime}$ delimited
by AT lines.

\vskip 2\baselineskip
\noindent
{\bf Fig. 3:} \qquad Phase diagram $T$ versus $J_{0}$ (in units of $J$) of
the SK model in the presence of a trimodal random field with $p_{0}=0.3$,
compared with one of the bimodal case ($p_{0}=0$), for conveniently
chosen values of $h_{0}$, in such a way as to obtain two tricritical points
(black circles) along the ferromagnetic boundary.
(a) $h_{0}/J=0.97$ \ ($p_{0}=0$);
(b) $h_{0}/J=1.558$ \ ($p_{0}=0.3)$.
The dashed lines stand for first-order phase transitions. The phase
nomenclature is the same as in Fig. 2.

\vskip 2\baselineskip
\noindent
{\bf Fig. 4:} \qquad Phase diagram $T$ versus $J_{0}$ (in units of $J$) of
the SK model in the presence of a trimodal random field with $p_{0}=0.3$,
compared with one of the bimodal case ($p_{0}=0$), for conveniently
chosen values of $h_{0}$, in such a way as to obtain a single
tricritical point (black circle) along the ferromagnetic boundary.
(a) $h_{0}/J=1.02$ \ ($p_{0}=0$);
(b) $h_{0}/J=1.58$ \ ($p_{0}=0.3$).
The phase nomenclature and line representations are as in Figs. 2 and 3.

\vskip 2\baselineskip
\noindent
{\bf Fig. 5:} \qquad The zero-temperature phase diagram $h_{0}$ versus
$J_{0}$ (in units of $J$) of the SK model in the presence of a trimodal
random field, for typical values of $p_{0}$. If $0<p_{0}<p_{0}^{*}$
one always gets tricritical points (black circles), followed by
first-order phase transitions for high values of $h_{0}$.
When $p_{0}=p_{0}^{*}$, one gets a fourth-order critical point (represented
by a star). Above the threshold value
$p_{0}^{*}=2(e^{3/2}+2)^{-1} \approx 0.30856$, the critical frontier
separating the phases {\bf SG} and {\bf F}$^{\prime}$ is continuous.

\vskip 2\baselineskip
\noindent
{\bf Fig. 6:} \qquad The ordinate represents either the coefficient
$A_{3}^{\prime}$ or $A_{5}^{\prime}$ \ [Eqs. (3.7b) and (3.7c), respectively]
along the ferromagnetic boundary, for $p_{0}=0.3$ \ ($h_{0}/J=1.53526$) \
(dot-dahed lines) and $p_{0}=p_{0}^{*}$ \
($h_{0}/J=\sqrt{3}$) \ (full lines), as a function of temperature. In the
former case, $A_{3}^{\prime}=0$ at $T/J \cong 0.25$ \ (with
$A_{5}^{\prime}<0$), whereas in the latter, $A_{3}^{\prime}=A_{5}^{\prime}=0$
at $T=0$.

\vskip 2\baselineskip
\noindent
{\bf Fig. 7:} \qquad Ranges of $p_{0}$ and $h_{0}/J$ associated with distinct
behaviors for the ferromagnetic boundary.
(a) First-order phase transitions at finite and zero temperatures, with a
single tricritical point at finite temperatures;
(b) Two tricritical points with a first-order phase transition for finite
temperatures;
(c) Continuous phase transitions.

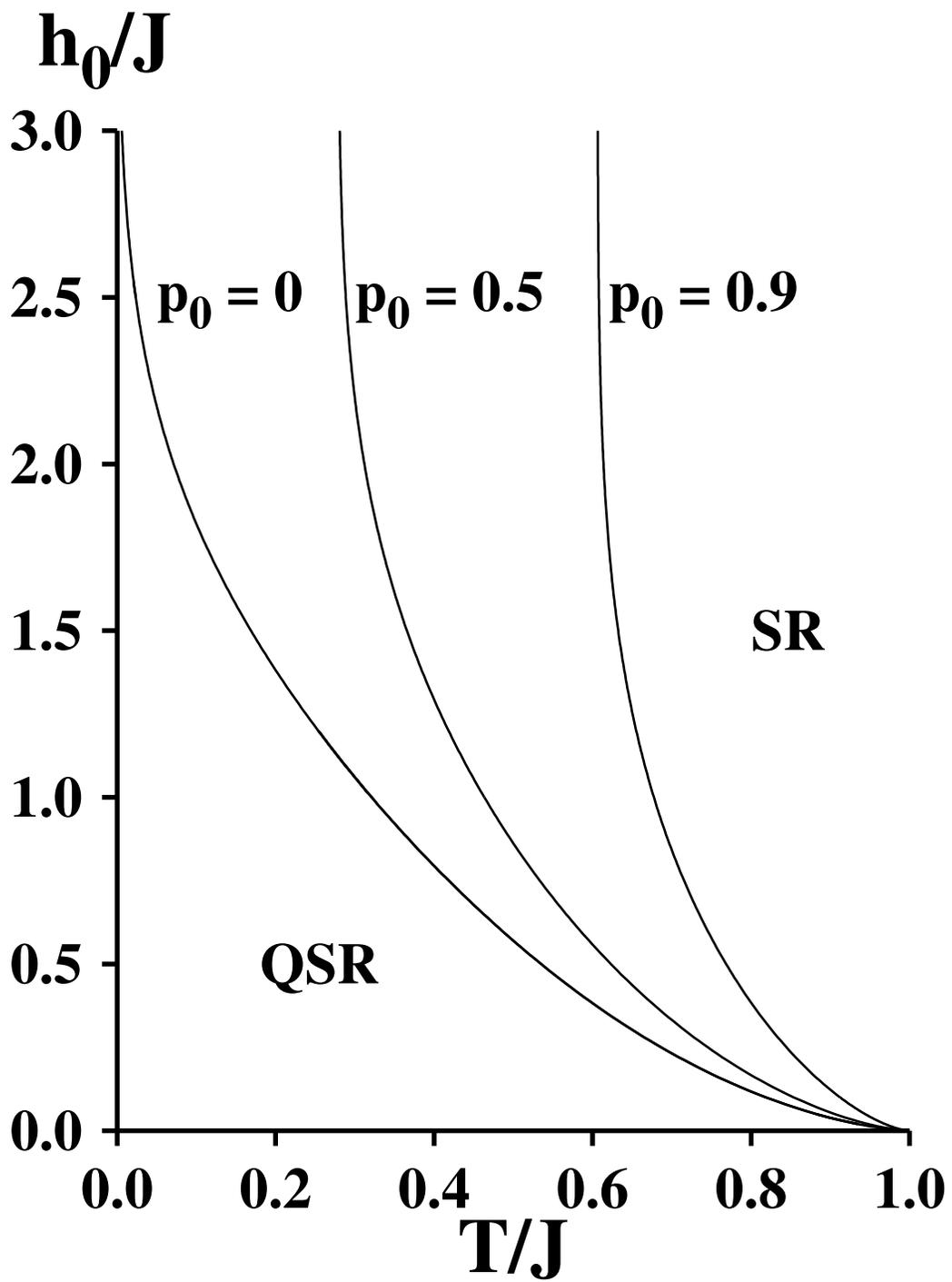
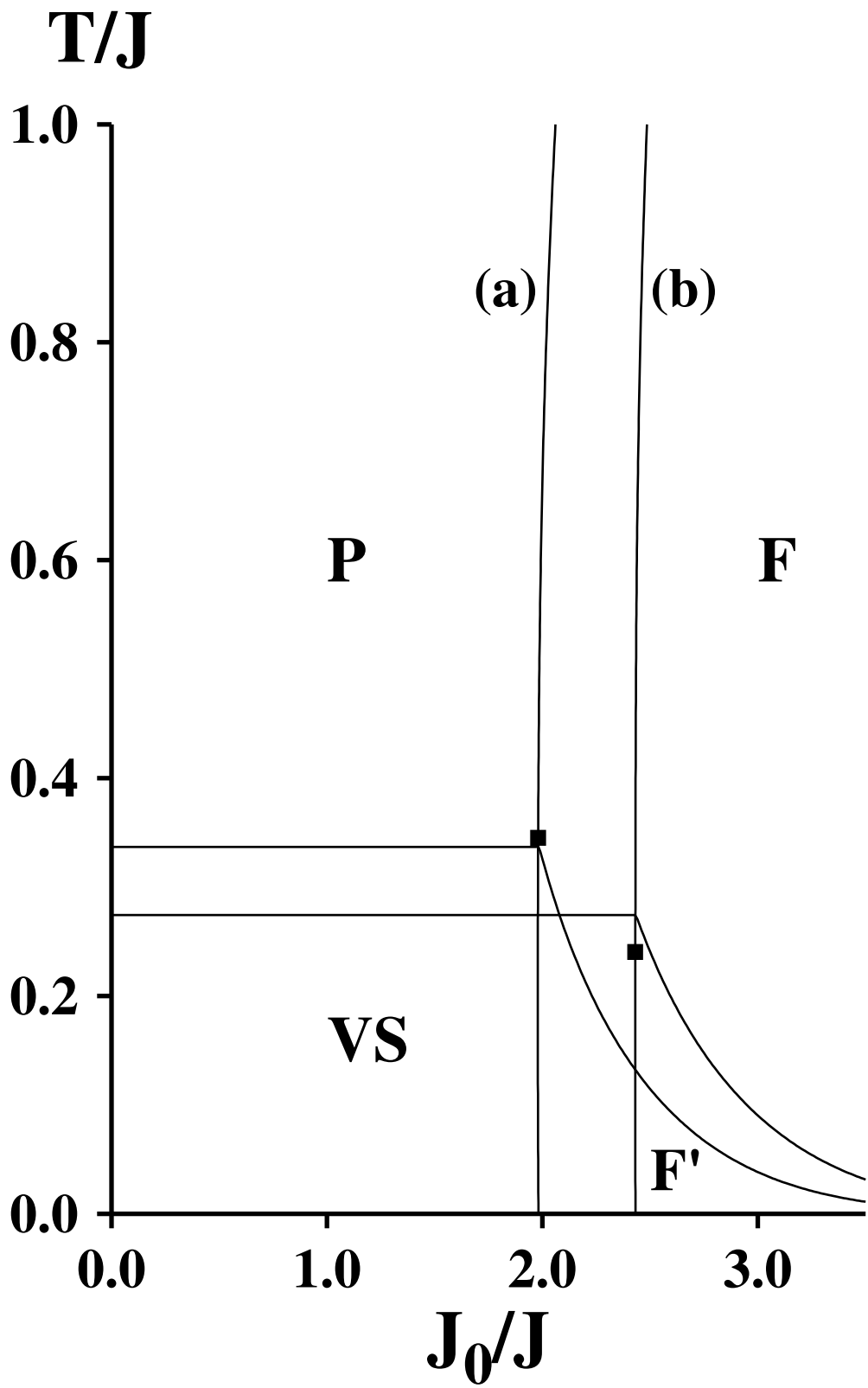
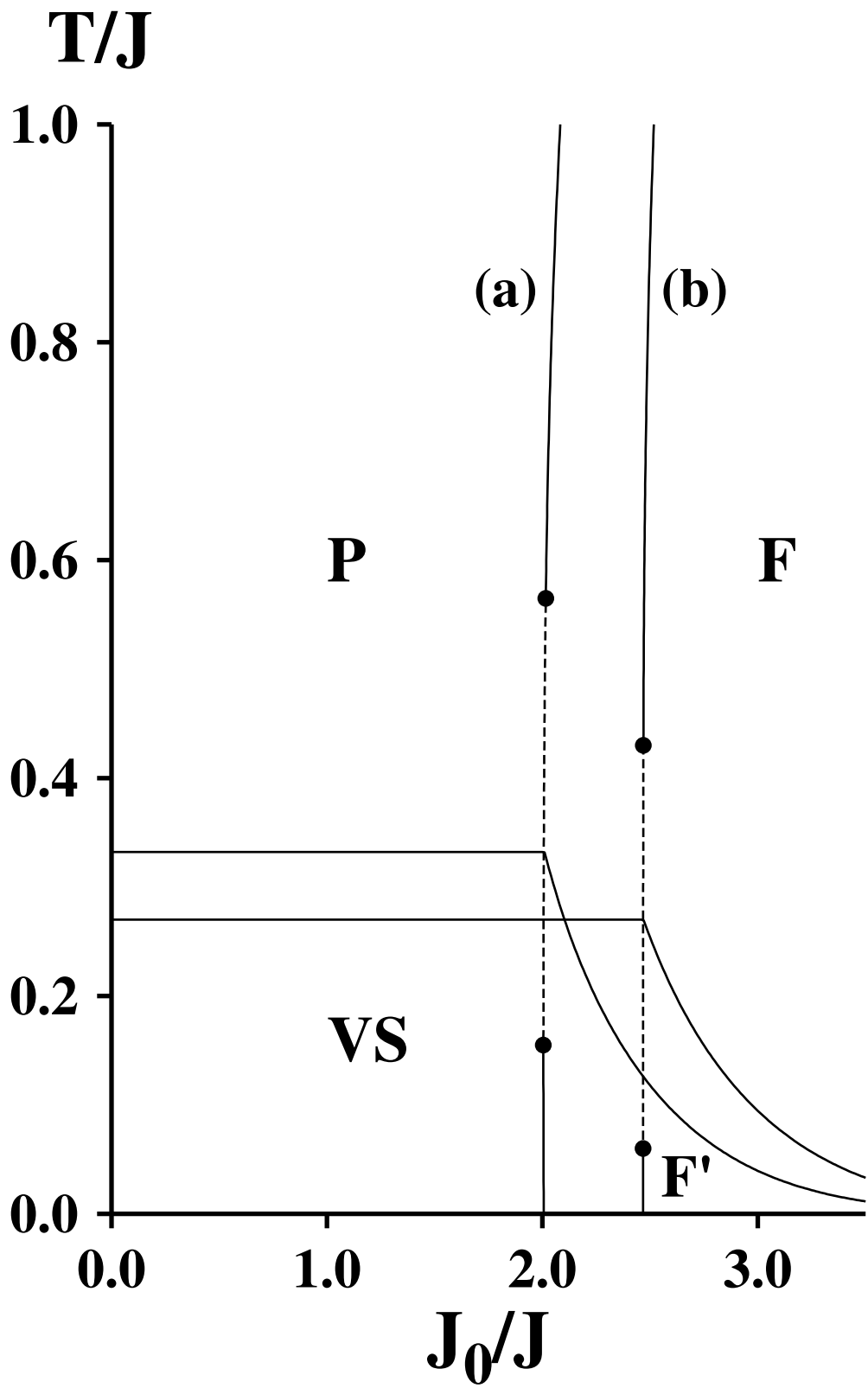
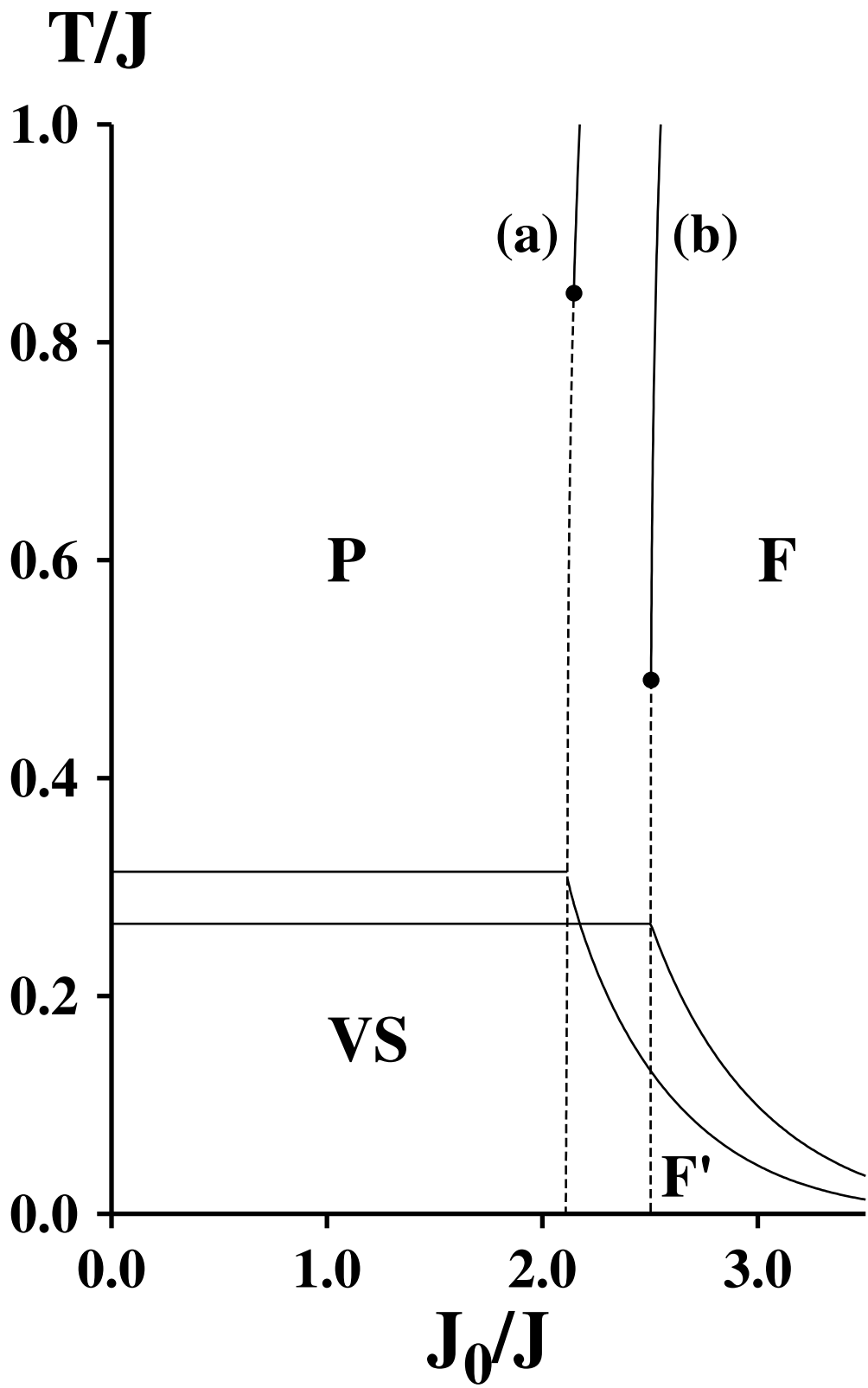
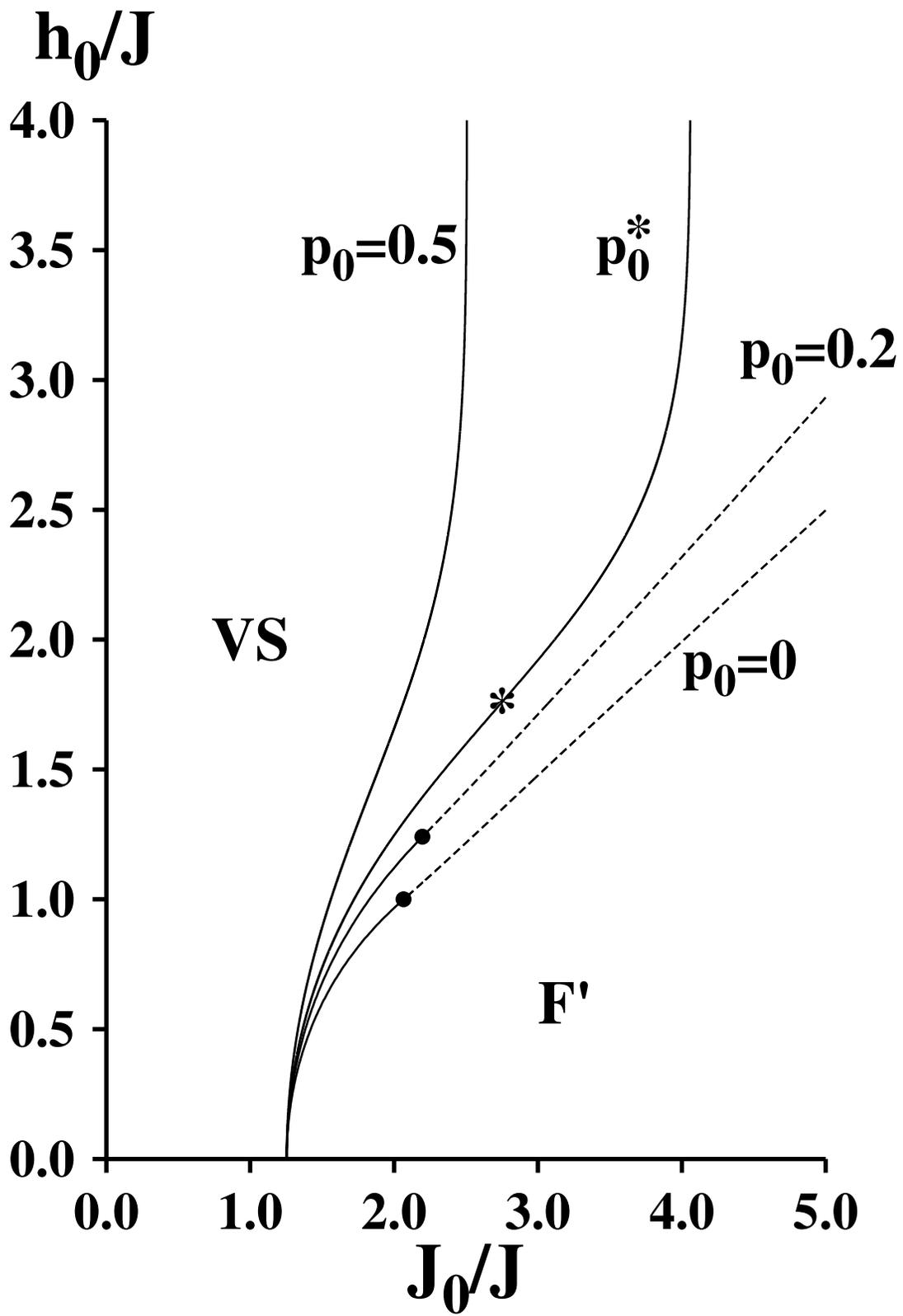
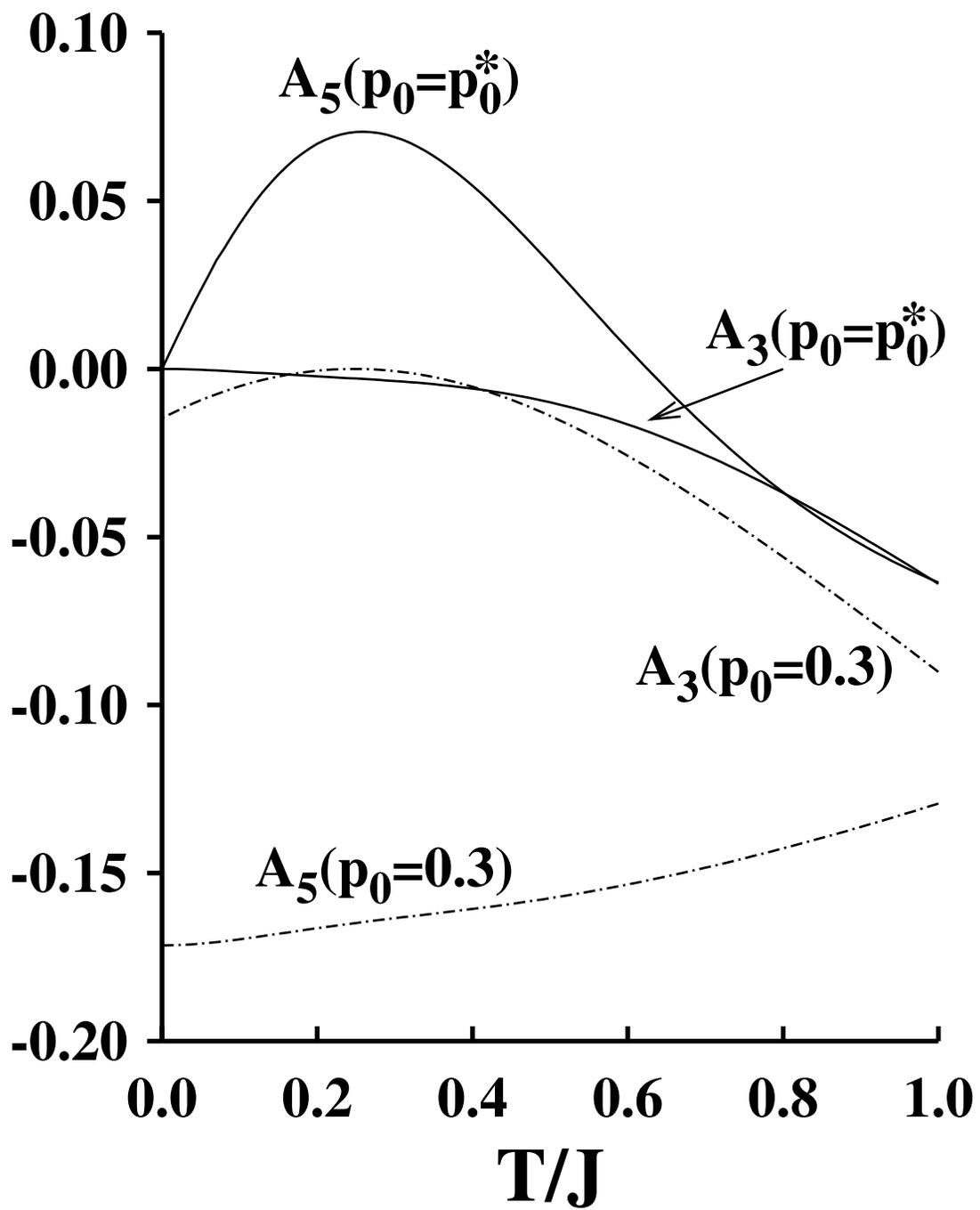
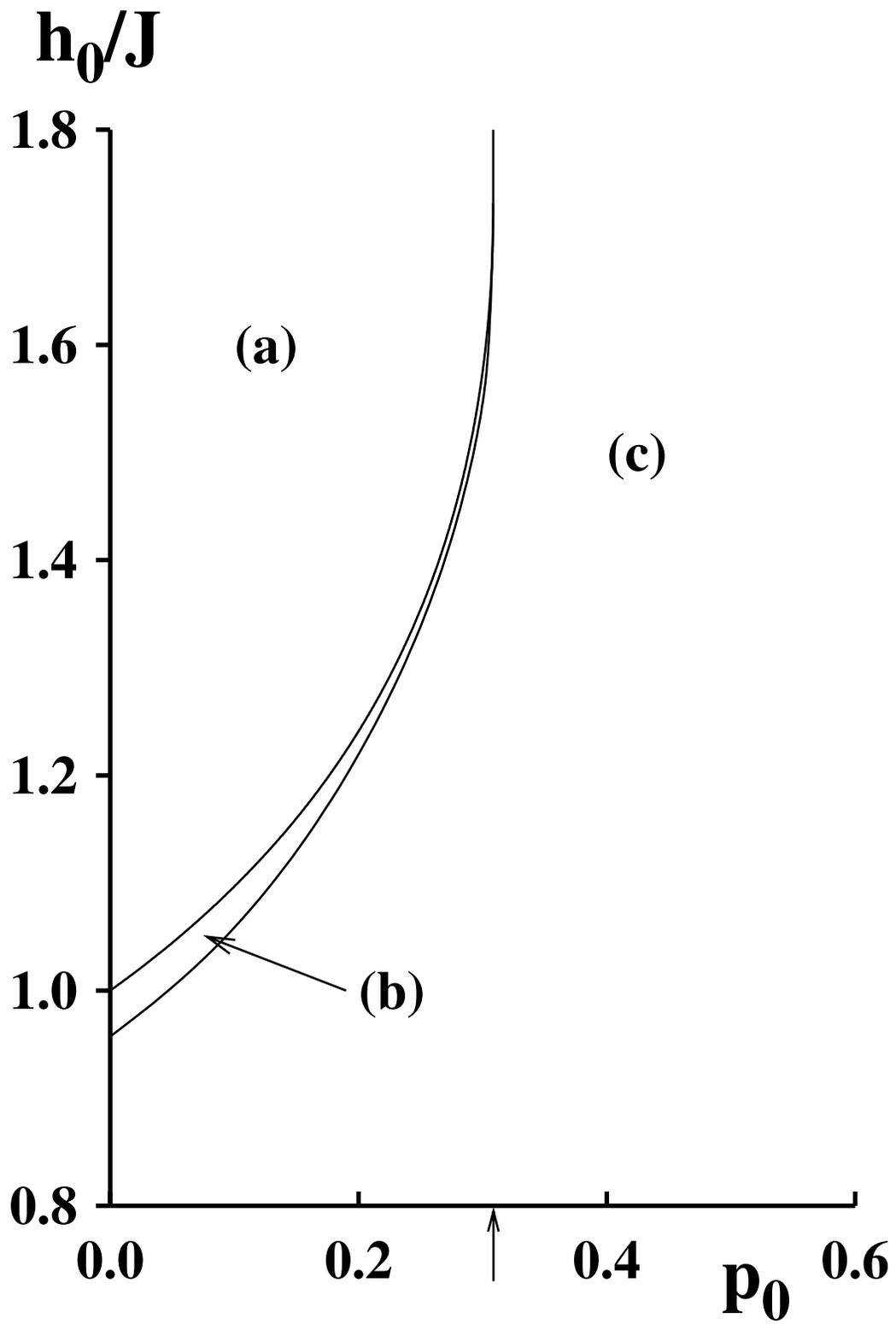
\end{document}